\documentclass[11pt]{article}
\usepackage[a4paper,left=25mm,right=20mm,top=20mm,bottom=20mm]{geometry}
\usepackage[utf8]{inputenc}
\usepackage{amsmath}
\usepackage{amsfonts}
\usepackage{amssymb}
\usepackage{graphicx}
\usepackage{epstopdf}
\usepackage{color}

\usepackage[english]{babel}
\usepackage{hyperref}

\title{Reconstruction of substrate's diffusion landscape by the wavelet analysis of single particle diffusion tracks}
\author{Eugene B. Postnikov\textit{$^{a}$} and Igor M. Sokolov\textit{$^{b}$}}
\date{
\textit{$^{a}$~Department of Theoretical Physics, Kursk State University, Radishcheva st., 33, 305000 Kursk, Russia. Fax: +7-4712-51-04-69; Tel: +7-4712-51-04-69; E-mail: postnicov@gmail.com}\\
\textit{$^{b}$~Institut f\"ur Physik and IRIS Adlershof, Humboldt Universit\"at zu Berlin, Newtonstra{\ss}e 15, 12489 Berlin, Germany. 
}
}

\begin{document}

\maketitle

\begin{abstract}
We propose an approach to analysing single trajectories of a particle, which moves randomly on a landscape distinct parts of which result in sufficiently various diffusion coefficients. The method based on the mapping the cumulative sum of step-wise elementary displacement squared into a complex oscillating functions with the subsequent continuous wavelet analysis of the later allows the localisation of relatively homogeneous sub-regions and determining values of the diffusion coefficient within each such sub-region. This approach is applied to demonstrable test examples as well as to the reconstruction of the biological membrane's properties from the real data of single molecule walk tracing.
\end{abstract}

%%%MAIN TEXT%%%%

\section{Introduction}
The method of single particle tracing is the method of choice for investigating diffusive properties of complex systems such as complex materials, biological cells or their membranes. 
What is of the major interest in many cases are not the properties of the diffusion itself, but its application as a probe for the properties of the environment \cite{Meroz2015,Manzo2015review}. 

Recently, several approaches to extract local properties of diffusive motion from single trajectories and to attributing them to specific spatially localized areas were proposed. 
In particular, one can mention clustering of diffusion parameters determined based on whole trajectory processing from a set of trajectories covering some spatial area \cite{Mellnik2014,Biermann2014}, 
segmentation of a single trajectory in parts corresponding to different diffusivities or into regions of active (superdiffusive) and passive (diffusive and subdiffusive) either by a sliding window analysis 
\cite{Alcor2009} or by the wavelet thresholding \cite{Chen2013}, enclosing parts of a single trajectory by convex hulls, geometrical properties of which distinguish between regions with different mobility rates \cite{Lanoiselee2017}, etc. 

However, in cellular environments the diffusion is not only often anomalous, being either sub- or superdiffusion \cite{Hofling2013, Sokolov2012,Meroz2015}, but can also be paradoxical. In this last case 
the width of the displacements' distribution grows linearly in time, but the distribution itself is strongly non-Gaussian. A theoretical example of extreme paradoxical diffusion 
with the distributions lacking moments is discussed in \cite{Sokolov1997}. In other, milder cases, the moments of the distribution (at least the second one) are present. This situation is termed as
``normal, yet non-Gaussian'' diffusion \cite{Forte2014,Chubynsky2014} see also \cite{Chechkin2017} and references therein. Such normal, yet non-Gaussian diffusion may have different reasons, and appear either due to the
heterogeneity of tracers, each possessing its own diffusion coefficient in an essentially homogeneous environment ("superstatistics"), due to environmental heterogeneity, or due
to fluctuations of the diffusion coefficients in time. To distinguish between such cases it is necessary to be able to distinguish between superstatistical and other cases, and to make 
a map of the local diffusivity maps of the environment with sufficient precision. The present work provides a method for performing this task. 

The situation assumed is as follows. A particle diffuses in an environment which might be homogeneous or not. The particle's local motion corresponds to an 
unbiased diffusion with the diffusion coefficient which might be a function of the position and time. These changes are considered gradual in space or slow enough in time, so that the 
fluctuations of the diffusion coefficient do not average out. A sufficiently long trajectory of the particle is recorded.
We aim at determining the value of the local diffusion coefficient for different parts of the trajectory. If many trajectories are recorded, each of them is processed in a similar way.
In the present work we concentrate on a two-dimensional case (membrane), for which the most data are available. 

The answers to our initial questions are given by analysing the fluctuations of the diffusion coefficient. If these are small, the total inhomogeneity has to be ascribed to the tracers'
differences. If these are ruled out, the next step would be analysing the difference of the diffusion coefficients on self-crossings of a single trajectory (which are multiple in two dimensions
due to the recurrent property of the corresponding random walks) or for two trajectories passing close by (in three dimensions). If the corresponding diffusion coefficients are close,
the medium can be considered as inhomogeneous. If they are different, the time-dependence of the diffusion coefficient due to the environment's rearrangement has to be assumed. 

\section{The method}
The input data of the SPT measurement are the time series of the instantaneous particle's positions $\mathbf{r}(t)$ sampled at the instants of time $n\Delta t$.
The total displacement from the beginning of the observation up to time $t$ corresponding to $N=t/\Delta t$ time steps can therefore be reprersented as a
displacement in a random walk:
\begin{equation}
 \mathbf{S}_N = \sum_{n=1}^N \mathbf{s}_n.
\label{disp}
\end{equation}
The squared displacement in the walk is 
\[
S^2_N = \left( \sum_{n=1}^N \mathbf{s}_n \right)^2 = \sum_{n=1}^N s^2_n + \sum_{n\neq n'} \mathbf{s}_n \mathbf{s}_{n'}. 
\]

Our random walk may be a correlated or not. If the random walk does not possess any memory at the timestep of the data aquisition procedure, the ensemble mean of 
each of the contributions to the second sum vanishes, and we get for the MSD
\[
R^2(t) = \langle S^2_N \rangle = \sum_{n=1}^N \langle s^2_n \rangle.
\]
If we assume the steps to be i.i.d. random variables possessing the zero first moment $\langle \mathbf{s} \rangle =0$ and a finite second moment $\langle s^2_n \rangle = \langle s^2 \rangle < \infty$, 
the sum will grow linearly in $N$ and therefore in time, and the slope of the dependence will define the diffusion coefficient $D=\langle s^2 \rangle/2 d \Delta t$ where $d$ is the dimension of space: 
for a simple random walk the diffusion coefficient is proportional to the mean squared displacement per step. The value $\langle s^2 \rangle$ can be estimated via time
averaging over $s_n^2$, or can be read out from the slope of the $R^2(t)$ curve of the integrated (summed) squared displacements, which is obtained by the linear fit to this dependence,
which is essentially the same. This can be expressed in the following words: neglecting the correlations and summing up squared displacements gives, for a single trajectory, the same 
result as it would be obtained when averaging over the thermal histories in many trajectories. 

Now let us imagine that the diffusion coefficient and therefore the mean squared step lengths vary in time, but the change in this coefficient is small on the timescale of a single step, 
so that $\langle s^2_n \rangle$ is a slowly changing function of $n$ or $t$. Then looking at $R^2(t)$ we still can read out $D(t)$ as the local slope of the corresponding curve. Of course the 
value of $D(t)$ can be also defined via the moving time average, which may even be theoretically preferable, but looking at the local slope has several practical advantages as will be discussed below.  

The same approach can be used for determining local diffusion coefficients in the case when the diffusion coefficient is position-dependent (but varies only slightly on the scales of a single step). 
In this case the local slope of the $R^2(t)$ curve gives us the diffusion coefficient at a position at which the particle is at time $t$. This also allows to distinguish between the time- and position-dependent
diffusion coefficients, at least in two dimensions, where the walks are recursive. Thus, if arriving at the same position the tracer shows a very different diffusion coefficient, the idea of 
purelly position-dependent diffusion coefficient (like in the static patchy environment) should be abandoned. Such situations will be discussed in the next Section by presenting
an example of simulated random walk in a heterogeneous environment.

One of the advantages of the method based on slope estimation is the fact that the approach based on moving time averaging of the local squared displacements 
poses a task of judicious choice of the averaging window, which, for achieving satisfactory accuracy, has to be chosen adaptively, depending on the local diffusion coefficient itself,
while the slope is easily grasped from the figure. Looking at the curves makes the behaviour evident for the naked eye, but for full analysis one still has to state on what scales the slope has to be estimated, 
where the change points are, etc. This can be made easy by using a robust approach to linear fitting connecting this with a wavelet transform, as will be discussed in Sec.~\ref{wavelets}.

Since at short times diffusion always wins over the deterministic motion, the estimate for the diffusion coefficient would not be bad even for a particle moving in some external potential, provided 
the interval $\Delta t$ between the data acquisition points is short enough. The only problem with estimating the diffusion coefficient from the local step lengths may occur when the steps of the random 
walks are strongly correlated. 
For short-ranged correlations the problem may be cured by comparing the $R(t)$ dependences at different $\Delta t$,
i.e. by coarse-graining or decimation procedures. In the case of the diffusion in a heterogeneous environment or when the properties of the process are explicitly time-dependent, 
it could be also nice to have ``early warning signs'' for the appearence of correlations in some patches or in some intervals of time. One of the approaches to getting such signs is discussed in Sec.\ref{Caveats}.

\section{Local linear fits by wavelet transforms}
\label{wavelets}

In the case of a random walk with independent and stationary individual steps, the problem of diffusivity identification
reduces to the problem of linear fit of the experimentally determined function $R^2(n)$ which might however show strong local fluctuations.
To understand the idea of a linear fit by a wavelet transform let $Y_n= R^2_n$ be equidistantly sampled data, with the total of $N \gg 1$ measurement points, 
and $y_n = kn+\beta$ be the best linear fit to it. 
Let us assume that $Y_n = k_0 n + \beta_0  + \xi_n$, and look at the distribution of the difference $\Delta_n = Y_n - y_n = Y_n -k n -\beta = (\beta_0-\beta) + (k_0-k)n + \xi_n$.
The value of $k$ only affects the position of the distribution of $\Delta$, the value of $\beta$ also affects the width. Introducing the shifted variable 
$\tilde{\Delta_n} = \Delta_n -(\beta_0-\beta) = (k_0-k)n + \xi_n$ we see that the distribution of this variable is a convolution of the distribution of $\xi_n$
and a rectangular distribution of width $(k_0-k)N$ centered at $N/2$. The properly defined width of the convolution of two distributions (i.e of the distribution of the 
sum of two independent random variables) is larger than the width of the distribution of each of them. 

In the standard least square fit (assuming the Gaussian statistics of errors, and an absence of the 
systematic error, i.e. that $\langle \xi_i \rangle = 0$) the width of the distribution is characterised by the mean squared value of $\tilde{\Delta}$, so that what 
has to be minimized is the sum of deviations squared. This definition makes problems when the statistics of $\xi_n$ is non-Gaussian: it is known that the least square fit is
not robust against outliers.

An alternative method \cite{Postnikov2015} is based on looking at the empirical characteristic function of $\tilde{\Delta}$
\[
 f(\Omega)= \langle e^{i \Omega \tilde{\Delta}} \rangle = \frac{1}{N} \sum_{i=1}^N \exp(i \Omega \tilde{\Delta}_n), 
\]
being an approximation to the Fourier transform of the probability density of $\tilde{\Delta}$. 
The narrower the distribution of $\tilde{\Delta}$ the broader is its spectrum $|f(\Omega)|$ (or $|f(\Omega)|^2$) and vice versa. 
The method is robust since the characteristic function exists for any distribution. 

For Gaussian distribution of errors, or for any other distribution of errors possessing the second moment $\sigma^2$ the form of this characteristic function
close to its maximum attained at $\Omega=0$ will be $f(\Omega) = 1+i\Omega (k_0-k)N/2-[\sigma^2 + \gamma^2 (k-k_0)^2] \Omega^2 + o(\Omega^2)$,
where $\gamma^2$ is the second moment of the distribution of the values of $n$, i.e. approximately $N^2/12$. Therefore the method is based
on fixing some $\Omega$ small enough (the criteria for a reasonable choice are discussed in Ref.\cite{Postnikov2015}) and looking for 
the maximum of the value of $|f(\Omega|k)|$ as a function of $k$ (taking the absolute value removes the influence of the term linear in $\Omega$ so that
for the Gaussian case the method is equivalent to the least square fit). We note that passing into the complex domain, and using the characteristic
function has additional advantages when considering the windowed (i.e. local) slopes, when the procedure can easily be represented as a wavelet transform.
We note moreover that the method still works when the distribution of errors lacks the second moment \cite{Postnikov2015}, when the least square fit fails.
Thus, to perform the linear fit one considers the function 
\begin{equation}
f(\Omega') = \left|\sum_n e^{i\left[\Omega R^2(n)-\Omega' n\frac{\mathrm{max}\left(R^2\right)}{N-1}\right]} \right|
\label{kdiff}
\end{equation}
where $\Omega$ is an appropriate scaling factor discussed above, and finds the value of $\Omega'$ corresponding to the maximum of this function.

The value $\Omega_m = \mathrm{argmax} f(\Omega')$ will give the desired diffusion coefficient 
\begin{equation}
D=(\Omega_m/\Omega)\cdot\left[\mathrm{max}\left(R^2\right)/((N-1)\Delta t)\right]/4.
\label{Dform}
\end{equation}
The prefactor of $\Omega'$ in Eq.(\ref{kdiff}) is chosen in such a way that for the case of linear $R^2(n)$ growing linearly in time $\Omega_m = \Omega$.

If $D(x,y)$ is position-dependent, the approach (\ref{kdiff}) needs to be restricted to the analysis of parts of random walk trajectories localized within relatively small spatial regions
where the diffusivity can be considered as approximately constant. This translates into the analysis of the parts of trajectories bounded to some time interval 
during which the displacement is relatively small. Such localisation can achieved by the replacement of the Fourier kernel by the Morlet wavelet with some central frequency $\Omega_0$
\begin{equation}
w(a,b)=\int\limits_{-\infty}^{+\infty}e^{i\Omega R^2(t)}e^{-i\Omega_0\frac{t-b}{a}}e^{-\frac{(t-b)^2}{2a^2}}\frac{dt}{\sqrt{2\pi a^2}},
\label{Morlet}
\end{equation}
(here $t$ stands for $n \Delta t$ and integral notation is used instead of the sum). This approach is similar to the method applied in \cite{Thiel2016} 
for the analysis of temporal evolution of non-stationary relaxation processes. 
The variable $b$ parametrizes the position of the part of the trajectory (in time) where the diffusion coefficient is estimated. 
The variable $a$ is called ``scale'' and plays a twofold role. First, it determines the duration of the
corresponding part of the trajectory, i.e. the width of the Gaussian window in which the diffusion coefficient is determined. 
Second, it is directly connected with the value of this local diffusion coefficient since the frequency $\Omega_m$ from Eq.~\ref{Dform} is equal to 
$\Omega_m=\Omega_0a^{-1}_{max}$, where $a_{max}$ is the scale corresponding to the maximum of the absolute value of the wavelet transform, $|w(a,b)|$ 
at a fixed $b$. 
Due to the properties of complex exponential, (\ref{Morlet}) can be rewritten as 
$$
w(\Omega'=\Omega_0a^{-1},b)=e^{-i\frac{t-b}{a}}\int\limits_{-\infty}^{+\infty}e^{i\left[\Omega R^2(t)-\Omega' t\right]}
e^{-\frac{(t-b)^2}{2(\Omega_0/\Omega')^2}}\frac{dt}{\sqrt{2\pi (\Omega_0/\Omega')^2}}.
$$
Note that the factor moved outside of the integral does not influence $|w(a,b)|$ and, therefore, can be further omitted while one considers the absolute value only. The integrand consists 
of the term coinciding with (\ref{kdiff}) multiplied by the bell-shaped Gaussian sliding filter, which assures time localization \cite{Postnikov2009,Postnikov2017} for the regions, 
where the residual $\Omega R^2(t)-\Omega' t$ tends to a constant. %, i.e. is minimally influenced by the multiscale Gaussian diffusive smoothing. 

The Morlet wavelet provides the best possible simultaneous localization in time and frequency domains and an exact correspondence between the maximum 
of its absolute value and the Fourier frequency of a harmonic signal, which in our case corresponds to the constant diffusion coefficient. 
The dark side of this property is that good frequency localization reveals a lot of collateral frequency components in the case of noisy signal. Moreover, 
aiming on obtaining the local diffusion coefficients we might stress the time localization over the frequency one, i.e. over the accuracy of the 
estimate for $D$.

In principle, the Gaussian function can be replaced by another well-localized filter of self-similar shape. To get better time localization mentioned above
the filters with finite support could be preferred. 
Thus, an appropriate replacement of (\ref{kdiff}) in the case of time-localized trajectory analysis will be 
\begin{equation}
K(\Omega',n)=\left|\sum_n e^{i\left[\Omega R^2(n')-\Omega' n'\frac{\mathrm{max}\left(R^2\right)}{N-1}\right]}
W\left[(n-n')\right]\right|.
\label{kdiffloc}
\end{equation}
A good candidate for the window function $W$ is a Tukey filter \cite{Harris1978} of length $M$ with $\alpha$: 
\begin{equation}
W(n)=
\left\{
\begin{array}{ll}
\frac{1}{2}\left[1+\cos\left(\pi\left(\frac{2n}{\alpha(M-1)}-1\right)\right)\right],&0\leq n\leq \frac{\alpha(M-1)}{2},\\
1,&\frac{\alpha(M-1)}{2}\leq n \leq (M-1)\left(1-\frac{\alpha}{2}\right),\\
\frac{1}{2}\left[1+\cos\left(\pi\left(\frac{2n}{\alpha(M-1)}-\frac{2}{\alpha}+1\right)\right)\right],&
(M-1)\left(1-\frac{\alpha}{2}\right)\leq n \leq(M-1).
\end{array}
\right.
\label{Tukey}
\end{equation}
It comprises a flat top, within of which (\ref{kdiffloc}) reduces to (\ref{kdiff}), and rapidly decaying side parts smoothly connected to the central one that assures locality of the 
filter and absence of significant spectral disturbances. In contrast to the conventional Windowed Fourier Transform \cite{Harris1978}, the length of the Tukey filter considered 
within a general multiscale theory of wavelets, should be coordinated with the characteristic periods corresponding to the values $\Omega_n$ at which $\Omega'$ is sampled, so that
$M=k \cdot 2\pi(N-1)/\Omega_n\mathrm{max}\left(R^2\right)$ rounded to a next integer. Here $k$ counts a number of periods $2\pi/\Omega_n$ stacked on the length of the filter. 
In further examples we use $k=3$ and $\alpha=0.5$, which provides a uniform averaging over two trial periods of oscillations with half-period-long smooth transition to zero value.
 The resulting local diffusion coefficient $D(n)$ corresponds to the frequency $\Omega_{m}(n)$, which maximizes (\ref{kdiffloc}) at the time point $n$ and again will be calculated using the formula (\ref{Dform}).

Before going into details of the procedures as applied to real data, let us discuss a small simulation of a random walk in a patchy environment, where the diffusion in each patch is normal
but is characterized by a different diffusion coefficient. 

\section{Numerical example}

To discuss the method let us consider a simple example of the discrete lattice random walk determined as 
\begin{align}
x(n+1)=&x(n)+s_x(n,\theta),\label{rwx}\\
y(n+1)=&y(n)+s_y(n,\theta),\label{rwy}
\end{align}
where $s_{x,y}$ is the step displacement. This displacement is either 0 or $\pm 1$ in one of the directions, and its value depends on the 
waiting time $\theta$ on a site (taken to be a whole number). If the walker arrives at a site at $n$-th time step, it stays there for the next $\theta$ subsequent time steps, 
and the next step in either direction is possible on the time step $n+\theta+1$. This determines the effective diffusivity as $D=2(1+\theta)^{-1}$. Translated to a continuous setting
the scheme of symmetric walks with site-dependent waiting times leads to the Ito interpretation of the corresponding stochastic differential equation \cite{Sokolov2010}.

\begin{figure}%
\includegraphics[width=\columnwidth]{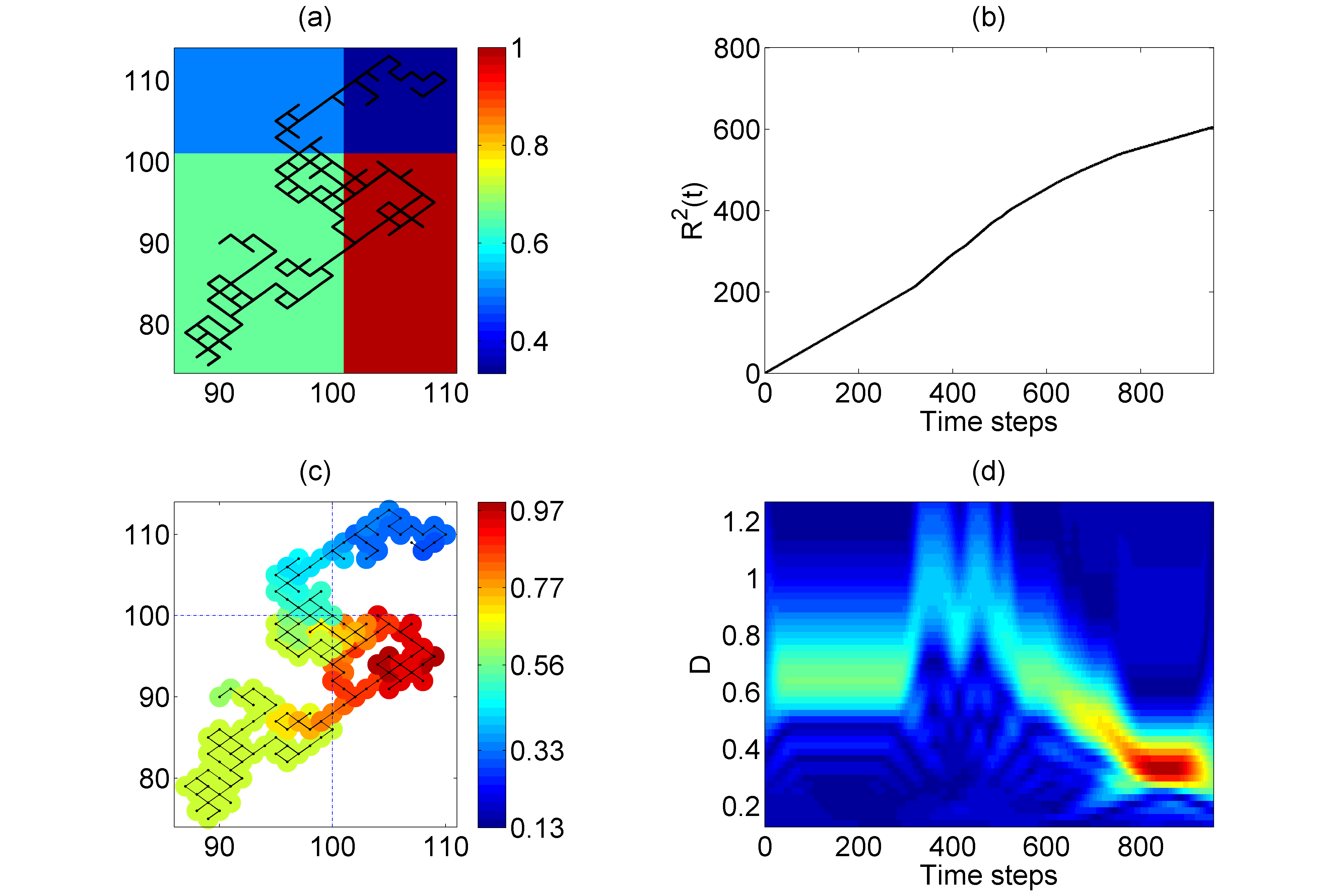}%
\caption{A trajectory (a), the cumulative displacement squared (b), the absolute value of the wavelet transform (d) and the reconstructed distribution of diffusivities along the trajectory (c)
for a numerical example discussed in the text.}%
\label{fig_waiting_subdivision}%
\end{figure}

As an example we consider a random walk generated according to Eqs.~(\ref{rwx})--(\ref{rwy}) on a plane subdivided into quadrants with the waiting times $\theta=1,2,3,5$ counter-clockwise starting from the right lower quadrant, see 
Fig.~\ref{fig_waiting_subdivision}(a). In this plot the coloured background indicates regions with different constant diffusion coefficients and the black line shows the random walk trajectory of 
951 time steps. Such relatively small length is chosen respectively to the characteristic lengths of samples, which can be obtained in realistic physical experiments \cite{Manzo2015review}. 
Fig.~\ref{fig_waiting_subdivision}(b) demonstrates the behaviour of $R^2(t)$, the sum of the step lengths squared, which will be substituted 
into Eq.~(\ref{kdiffloc}), resulting in abdolute values of the wavelet transform as color-coded in Fig.~\ref{fig_waiting_subdivision}(d). The ordinate $D'$ 
is scaled from the parameter $\Omega'=a^{-1}\Omega_0$ in the same way as Eq.(\ref{Dform}) defines it for  the case of wavelet maximum $\Omega'=\Omega_m$. 
The corresponding estimates for $D$ correspond to a stripe of the brightest colour in the plot of the absolute value of the transform.

One can see that the relatively long first part of trajectory corresponding to the random walk on the quadrant with $D=0.67$ results in the constant slope subinterval of around 300 steps
in Fig.~\ref{fig_waiting_subdivision}(b), which is mapped into the horizontal bright stripe the left side of in Fig.~\ref{fig_waiting_subdivision}(d) of the same duration. The position
of the maximum returns to the same localization with after short excursions into the regions with larger diffusivity (indicated by maxima located higher in Fig.~\ref{fig_waiting_subdivision}(d)). 
After this, the walker moves inside the region of the lowest diffisivity (dark-blue background in Fig.~\ref{fig_waiting_subdivision}(a)) corresponding to the last stable horizontal 
subinterval in Fig.~\ref{fig_waiting_subdivision}(d). 

Using Fig.~\ref{fig_waiting_subdivision}(d) we determined the $D$, the position of the maximum of the absolute value of our wavelet transform 
(which in the following will simply be called the ``wavelet maximum'' for the sake of brevity) for each time step and represented the corresponding value of diffusivity 
on the walker's trajectory by a color code, similar to the one used to represent waiting times, see Fig.~\ref{fig_waiting_subdivision}(c). 
The corresponding colour coding clearly indicates four patches (red, yellow, neon blue, and dark blue), 
whose locations correspond to four quadrants of the original diffusivity distributions up to the some uncertainties at the boundaries of these sub-regions, 
where the boundary effects originated from a finite width on the Tukey window (\ref{Tukey}) occur. 

To demonstrate that the algorithm described above actually reconstructs the spatial patches of the diffusion coefficient and allows for distinguishing them from
possible time-dependent diffusivities, we consider another simple simulation, when the
waiting times governing $\delta_{x,y}$ in Eqs.~(\ref{rwx})--(\ref{rwy}) are assumed to be random variables chosen at arrival to the 
corresponding site and independent on the prehistory.
In the further example, $\theta(x,y)$ are taken as integers obtained by rounding of numbers uniformly distributed over the interval
$[1\,\theta_{max}]$ and, therefore
the averaged instant  displacement squared determines the effective averaged diffusivity as $D=2(1+\langle \theta\rangle)^{-1}$ with 
$\langle s\rangle=(\theta_{max}-1)/2$.

\begin{figure}%
\includegraphics[width=\columnwidth]{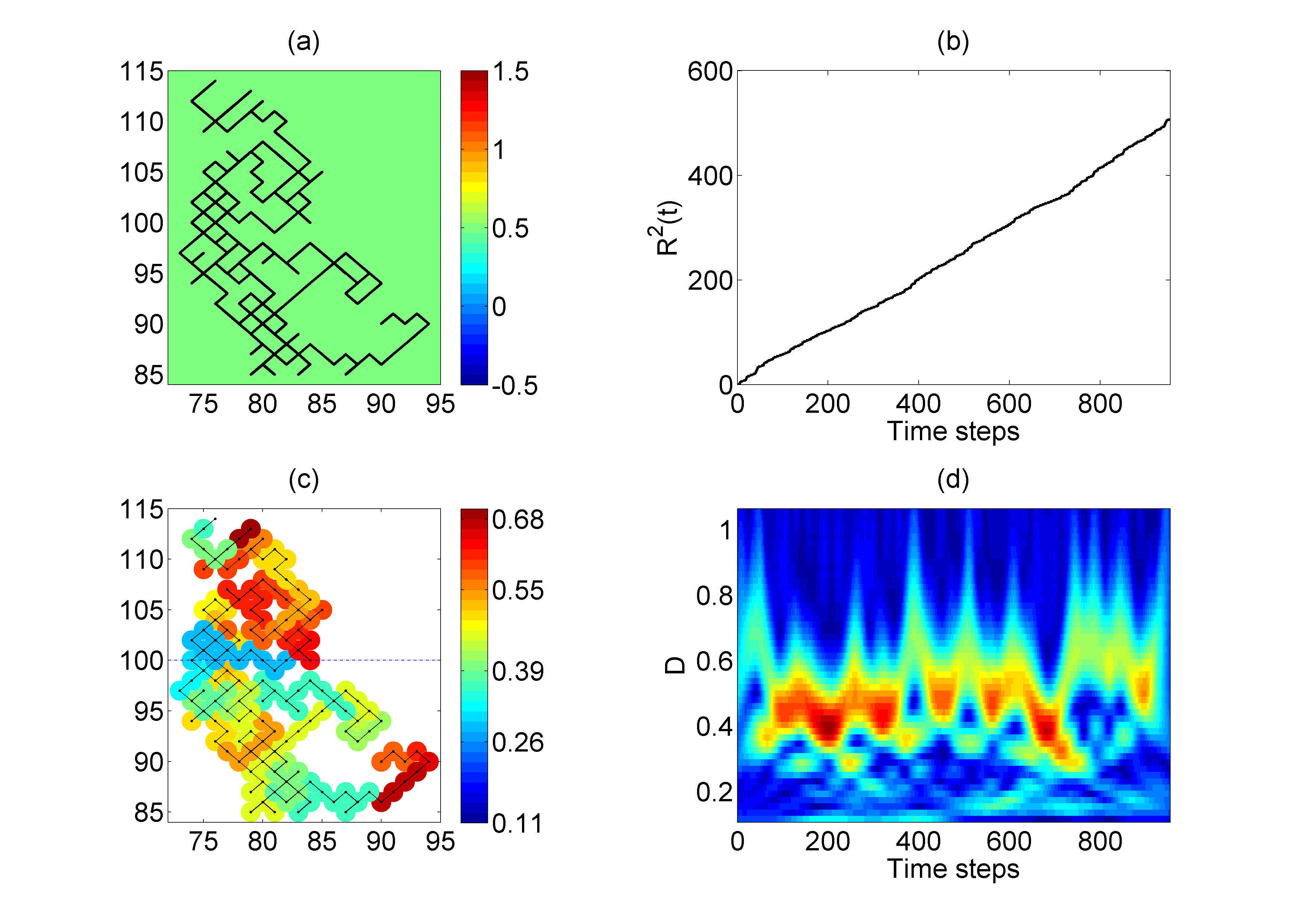}%
\caption{The same as in Fig.1 but for the case of the time-dependent diffusivity.}%
\label{fig_t_subdivision}%
\end{figure}

Fig.~\ref{fig_t_subdivision} illustrates such process for $\theta_{max}=5$. From Fig.~\ref{fig_t_subdivision}(d) one readily infers that the maxima of the 
absolute value of the wavelet transform are located in a stripe around $D=0.5$ corresponding to $\langle \theta\rangle=3$ that is close to the average slope of the line in
Fig.~\ref{fig_t_subdivision}(b). However, this stripe shows bursts, which reflect jumps in diffusivities. These bursts are
well-localized, i.e. they indicate fast changes of diffusivities localized in time. 

An additional representation of the trajectory color-coded for the local value of the diffusion coefficient, see Fig.~\ref{fig_t_subdivision}(c)  
unveils the effect which now is evident from the fact that the overlapping parts of the trajectory are coded with different colors: in contrast to the case of 
spatially-dependent diffusion considered earlier (Fig.~\ref{fig_waiting_subdivision}(c)) the parts of the trajectory going through the same coordinate region
are colored differently.

\section{Application to real data}

\begin{figure}
\includegraphics[width=\columnwidth]{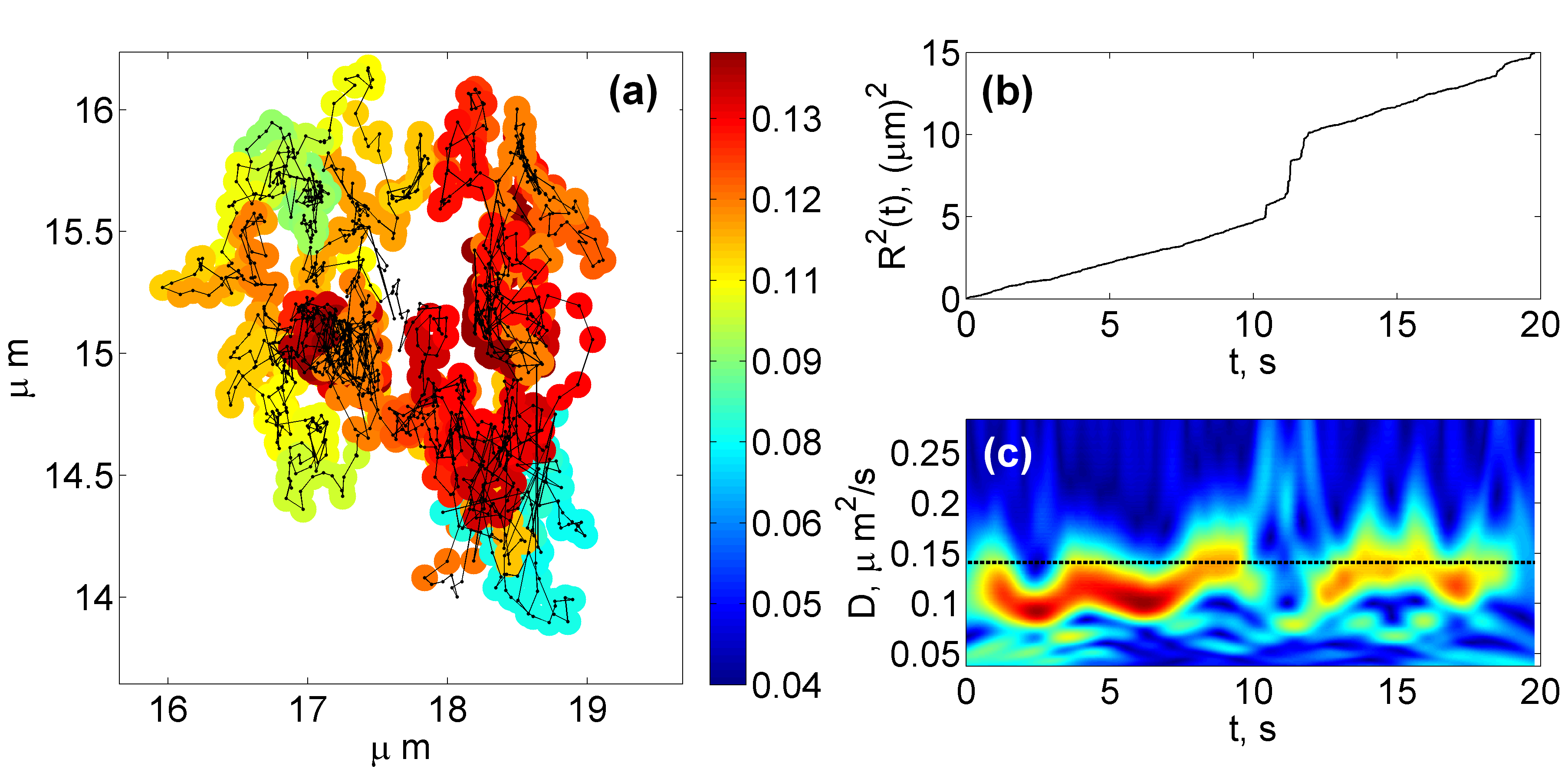}
\caption{(a) A trajectory of the motion of DC-SIGN protein on a living-cell membrane investigated experimentally in in the work \cite{Manzo2015} (the
data provided by courtesy of C. Manzo) shown in black, supplied with the values of local diffusion coefficient determined by
the wavelet-based method indicated by colour spots. (b) The cumulative displacement squared obtained from the trajectory. (c) The absolute value of 
the wavelet transform for $R^2(t)$, color-coded, see text for details.}
\label{figCarlo}
\end{figure}

As a practical example of the proposed algorithm application to real data of single molecule walk on a biological membrane, we analyse one of the data
sets measured experimentally and presented in the work \cite{Manzo2015} (the data provided by courtesy of C. Manzo), see
\ref{figCarlo}(a), where actual walker locations are marked by black dots connected by lines for visibility.
The plot $R^2(t)$ shown in Fig.~\ref{figCarlo}(b) is quite far from a straight line. However, there are sufficiently long intervals 
(almost a half of the track in total), where it can be approximated by such with high accuracy. Thus, one
can hypothesize that such behaviour originates from different diffusion coefficients corresponding to different properties of the membrane 
at different locations.

The result of application of Eq.~(\ref{kdiffloc}) with (\ref{Tukey}) with the function $R^2(t)$ depicted in \ref{figCarlo}(b) is shown in
Fig.~\ref{figCarlo}(c) that allows for the direct comparison of the behaviors of $R^2(t)$ and of the wavelet's maximum. Here warmer
colours correspond to larger values of the wavelet maximum. 

The dashed black straight line is the global average diffusion coefficient determined by the time-averaged mean-square displacement
\begin{equation}
tMSD(\tau=m\Delta t)=\frac{1}{N-m}\sum_{i=1}^{N-m}\left[\left(x(t_i+m\Delta t)-x(t_i)\right)^2+\left(y(t_i+m\Delta t)-y(t_i)\right)^2 \right]
\label{tMSD}
\end{equation}
with $\mathrm{max}(m)=18$ ($\mathrm{max}{\tau}=0.03~s$) equal to $D_{tMSD}=0.1309~\mathrm{\mu m^2\cdot s^{-1}}$ that agrees with the result given in \cite{Manzo2015}. 
The local maxima of the wavelet transform are close to the $tMSD$-based value of diffusivity, fluctuating around this line but also show bursts typical for rapid changes.

Fig.~\ref{figCarlo}(a) also presents the trajectory of this random walk color-coded for the local diffusivity obtained via the our wavelet maxima (maxima with
magnitudes less than 40\% of the largest value of the wavelet maxima  are excluded from the analysis since they correspond to 
finite size effects. The
resulting spatial picture can be read in a way discussed for the examples above: there are two regions of slow motions (yellow-green and cyan) separated by the region
of fast motion (red). The diffusivity in all these regions slightly fluctuates in time.

\section{Caveats and safeguarding procedures}
\label{Caveats}

As it has been mentioned above, the proposed procedure will result in an accurate representation of the local diffusion coefficients if the underlying
random displacements are indeed diffusive i.e. not correlated on the time scale of the step of data acquisition procedure. 
As a safeguarding procedure one can check whether this is indeed the case. This can either be done by decimation of the data, or by their local smoothing
before calculating the cumulative squared displacement $R^2(t)$. The local smoothing procedure has to be performed so that in the case of uncorrelated data
$R^2(t)$ would not change, but the changes are visible for correlated one. Such a procedure can be given a slightly different flavor, as we will discuss below.

We consider the localized smoothing of the random walk data based on the local multiscale averaging, each iteration of which is 
computed in the discrete case
\begin{equation}
s^{(j+1)}_n=\frac{s^{(j)}_{2n}+s^{(j)}_{2n+1}}{\sqrt{2}}, 
\label{sav}  
\end{equation}
for each component of the vector of elementary steps $\mathbf{s}_n=(s_x,s_y)$ (indices $x,\,y$ are omitted in Eq.~(\ref{sav})), which correspond to $j=0$.  
 
Thus, let us analyse now a set of the coarse-grained cumulative displacements squared
$$
R_{(j)}^2(t^{(j)}_N) = 2^j\sum_{n=1}^N  {s^{(j)}_n}^2,
$$
where the factor $2^j$ is introduced to compensate scaling in Eq.~(\ref{sav}), and time steps with respect to initial ones are defined as
$t^{(j+1)}_n=t^{(j)}_{2n}$, i.e. via power two of the elementary time steps.

In such representation, the function $R_{(j)}^2$ is invariant on the average for the true Brownian motion with uncorrelated steps and the
wavelet-based algorithm gives the same value of the local diffusion coefficients that follows from the multiscale construction of the Brownian motion using the sequential downward refinement within the
Haar wavelet basis \cite{Grebenkov2015}.
On the contrary, if the random walk is correlated, the invariance of $R_{(j)}^2$ is broken and the local diffusion
coefficients will differ for different levels of coarse graining $j$.

\begin{figure}%
\includegraphics[width=\columnwidth]{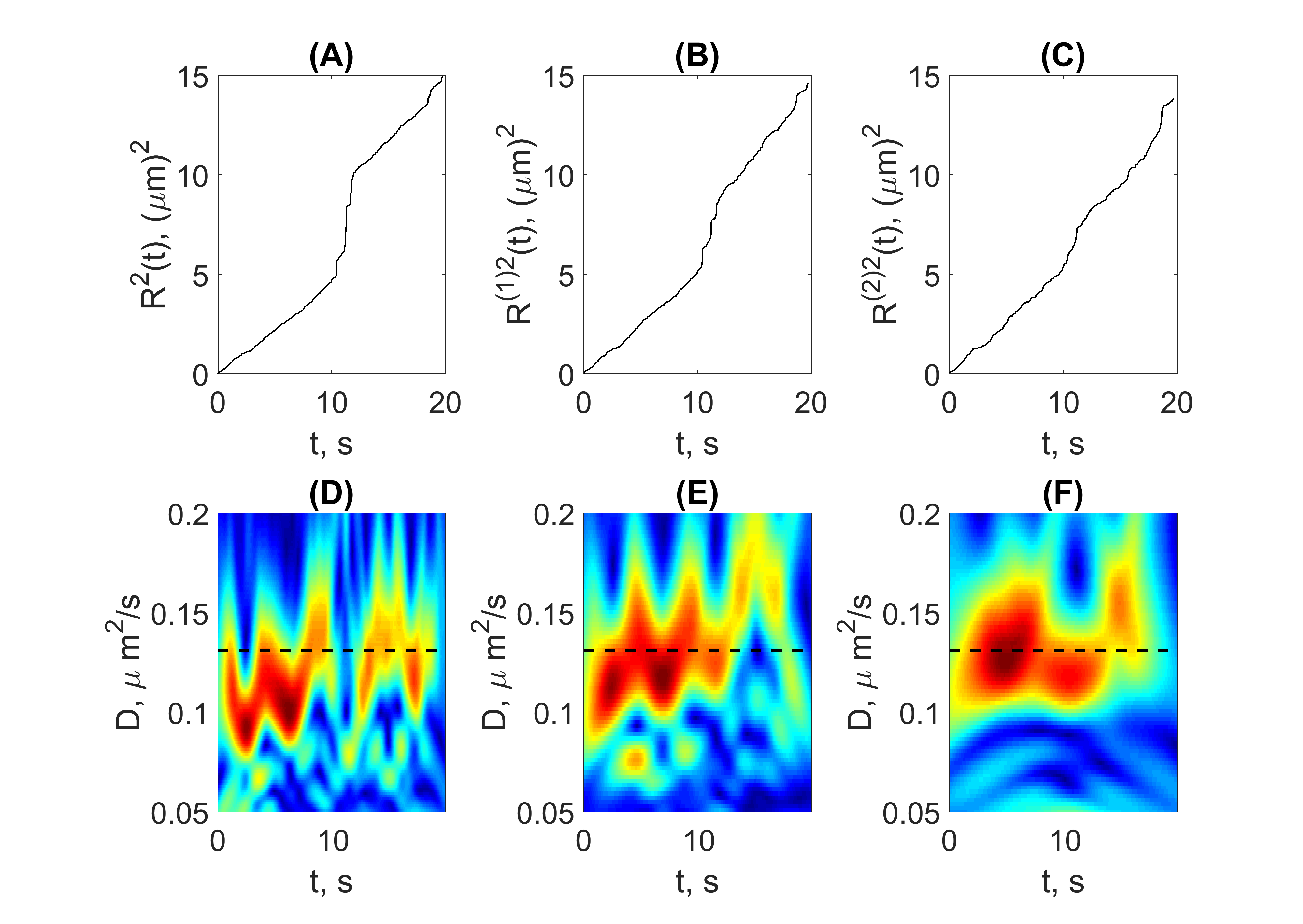}%
\caption{Cumulative local displacement squared for the initial random walk (A) and its two sequential coarse grainings (B)--(C). Panls (D)--(F) show
the corresponding plots of the wavelet maxima placed under respective cumulative trajectories: ordinates of brightest spots corespond to the
local diffusion coefficients for each time moment. The dashed line shows the average diffusion coefficient for the whole trajectory obtained by
(\ref{tMSD}).}%
\label{smooth}%
\end{figure}

Fig.~\ref{smooth} shows an illustrative example of such coarse-graining applied to the motion of DC-SIGN protein on a living-cell membrane
discussed above. One can see that the first and the last thirds of the cumulative trajectory $R^2$ are only weakly affected by the binary smoothing procedure
that indicates that the random motion on this subintervals is quite close to the Brownian. Certainly, Figs.~\ref{smooth}(E),~(F) for coarse grained signal have worse time resolution 
than the initial one Fig.~\ref{smooth}(D), but still the local diffusion coefficients obtained for the corresponding time domains and positions may be claimed true reasonable. 
On the other hand, the middle part, which contains combination of slow motion and long jumps is demonstrably affected by the smoothing procedure: the wavelet maximum within this subinterval 
moves upward during the smoothing. However, also in smoothed data the wavelet maxima lie in 
a stripe close to the value of the diffusion coefficient obtained by the standard time-averaged mean-square displacement
method.

Thus, the simple pair-wise averaging procedure with the subsequent application of the wavelet-based method provides facts of evidence for the
regions, where it give the true diffusion coefficient of the spatial- or time-dependent random walk, which need to be considered, or, as a byproduct,
indicates region of time-correlated random walk supplied with the estimation of characteristic correlation times.

\section{Conclusions}

The principal results of this work can be summarised as follows. Modern developments in single particle tracking not only open new perspectives for
the study of molecular motions in complex environments, but also allow for using these motions as a probe for revealing properties of substrates on
which the motions occur. This approach is especially important for studying biological membranes since their non-uniform structure (e.g. due to the
presence of protein clusters) results in variation of the local diffusion propertuies, as well as to a large variety of anomalous diffusion phenomena. 
The first step on this way is providing a map of local diffusion coefficients. Using a standard approach based on moving time averaging of the local 
squared displacements poses a task of judicious choice of the averaging window, which, for achieving satisfactory accuracy, has to be chosen adaptively, 
depending on the local diffusion coefficient itself, and is hardly practicable. We show that the method based on the wavelet transform of the cumulative squared
displacement assures for a better spatio-temporal localization. Its applicability was checked in numerical simulations of random walks on patchy 
substrates with different diffusion coefficients within patches as well as in
application to real experimental data on the single molecule tracking. We also not that mapping local diffusivities allows for 
distinguising diffusion on patchy substrates from the cases of random walks time-dependent diffusion coefficient which are another popular model for 
effects observed in biological systems.

\section*{Acknowledgment}

EBP is supported by the Russian Science Foundation, project 16-15-10252.

%%%REFERENCES%%%

%\bibliography{difbib} %You need to replace "rsc" on this line with the name of your .bib file
%\bibliographystyle{vancouver}

\end{document}